\documentclass[12pt]{article} 
\usepackage[sectionbib]{natbib}
\usepackage{array,epsfig,epstopdf, fancyheadings,rotating}
\usepackage[]{hyperref}  
\usepackage{sectsty, secdot}
\sectionfont{\fontsize{12}{14pt plus.8pt minus .6pt}\selectfont}
\renewcommand{\theequation}{\thesection\arabic{equation}}
\subsectionfont{\fontsize{12}{14pt plus.8pt minus .6pt}\selectfont}

\usepackage{amsmath}
\usepackage{amssymb}
\usepackage{amsfonts}
\usepackage{multirow}
\usepackage{amsthm}
\usepackage{color}

\newtheorem{theorem}{Theorem}
\newtheorem{lemma}{Lemma}
\newtheorem{corollary}{Corollary}

\theoremstyle{definition}

\newtheorem{example}{Example}

\newcommand{\ba}{\begin{eqnarray}}
\newcommand{\ea}{\end{eqnarray}}
\newcommand{\bas}{\begin{eqnarray*}}
\newcommand{\eas}{\end{eqnarray*}}

\newcommand{\e}{ { \mathbb{E}}}
\newcommand{\var}{ {\mathbb{V}\rm ar }}
\newcommand{\cov}{ {\mathbb{C}\rm ov}}

\newcommand{\ben}{\begin{enumerate}}
\newcommand{\een}{\end{enumerate}}

\def\T{{ \mathrm{\scriptscriptstyle \top} }}

\begin{document}


\renewcommand{\baselinestretch}{1.5}

\markboth{\hfill{\footnotesize\rm Chunlin Wang, Pengfei Li, Yukun Liu, Xiao-Hua Zhou, Jing Qin} \hfill}
{\hfill {\footnotesize\rm } \hfill}

\renewcommand{\thefootnote}{}
$\ $\par


\fontsize{10}{12pt plus.6pt minus .4pt}\selectfont \vspace{0.2pc}
\centerline{\normalsize\bf HYPOTHESIS TEST ON A MIXTURE FORWARD--INCUBATION-TIME}
\vspace{2pt}
\centerline{\normalsize\bf EPIDEMIC MODEL WITH APPLICATION TO COVID-19 OUTBREAK}
\vspace{2pt}
\centerline{\large\bf }
\vspace{.2cm} \centerline{
Chunlin Wang$^{1}$, Pengfei Li$^2$, Yukun Liu$^3$, Xiao-Hua Zhou$^4$, Jing Qin$^5$}
\vspace{.2cm}
\centerline{\it
$^1$Xiamen University, $^2$University of Waterloo, $^3$East China Normal University, }
\centerline{\it
$^4$Peking University, $^5$National Institutes of Health}
\vspace{.5cm} \fontsize{9}{11.5pt plus.8pt minus .6pt}\selectfont


\begin{quotation}
\noindent {\it Abstract:}
The distribution of the incubation period of the novel coronavirus disease that emerged in 2019 (COVID-19) has crucial clinical implications for understanding this disease and devising effective disease-control measures.
\cite{Qin2020} 
designed a cross-sectional and forward follow-up study to collect the duration times between a specific observation time and the onset of COVID-19 symptoms for a number of individuals. They further proposed a mixture forward--incubation-time epidemic model, which is a mixture of an incubation-period distribution and a forward time distribution, to model the collected duration times and to estimate the incubation-period distribution of COVID-19. In this paper, we provide sufficient conditions for the identifiability of the unknown parameters in the mixture forward--incubation-time epidemic model when the incubation period follows a  two-parameter distribution. Under the same setup, we propose a likelihood ratio test (LRT) for testing the null hypothesis that the mixture forward--incubation-time epidemic model is a homogeneous exponential distribution. The testing problem is non-regular because a nuisance parameter is present only under the alternative. We establish the limiting distribution of the LRT and identify an explicit representation for it.
The limiting distribution of the LRT under a sequence of local alternatives is also obtained. Our simulation results indicate that the LRT has desirable type I errors and powers, and we analyze {a COVID-19 outbreak dataset from} China to illustrate the usefulness of the LRT.

\vspace{9pt}
\par
\end{quotation}\par

\def\thefigure{\arabic{figure}}
\def\thetable{\arabic{table}}

\renewcommand{\theequation}{\thesection.\arabic{equation}}

\fontsize{12}{14pt plus.8pt minus .6pt}\selectfont

\setcounter{section}{0} 
\setcounter{equation}{0} 

\lhead[\footnotesize\thepage]{}\rhead[]{\footnotesize\thepage}%

\section{Introduction}
\label{sec1}

As the novel coronavirus disease that emerged in 2019 (COVID-19) spread rapidly worldwide, the World Health Organization (WHO) declared the COVID-19 outbreak a global pandemic on March~10, 2020. Currently, COVID-19 is still spreading around the world, posing a huge threat to global public health and having a huge impact on global economics and social development. As of January~7, 2022, the WHO had identified over 300 million confirmed cases of COVID-19 and observed more than 5 million deaths. Countries around the world have made great efforts to fight this pandemic by imposing various measures, such as isolation policies, travel restrictions, lockdowns, and social distancing. Among these measures, quarantining people who may have been exposed to COVID-19 seems to be the most effective way of preventing further disease transmission.

The incubation period of an infectious disease is the time between exposure to it and the first appearance of symptoms. Accurate estimation of the incubation-period distribution, or incubation distribution, is crucial (especially in regions where the epidemic is severe) for determining the length of appropriate quarantine periods for suspected individuals. In the literature, estimating incubation distributions has attracted much attention \citep{Sartwell1950, Kalbfleisch1989, Struthers1989,Kalbfleisch1991,Farewell2005, Wilkening2008}, while studies for COVID-19 are still ongoing; see \cite{Backer2020}, \cite{Guan2020}, \cite{Lauer2020}, \cite{Li2020}, \cite{Linton2020}, \cite{LiuXH2020}, \cite{Qin2020}, \cite{Rahman2020}, \cite{WangLili2020}, and \cite{LiuXH2021}, among others. The current results are based mostly on clinical experience or empirical statistical analysis of contact-tracing data, but such data may be inaccurate because of the patient's recall bias or the interviewer's personal judgment on the possible date of exposure rather than the actual date. More discussions can be found in  \cite{Qin2020}.

The lockdown of Wuhan, the capital city of Hubei province in China, provided an opportunity to estimate accurately the incubation distribution of COVID-19. \cite{Qin2020} designed a new cross-sectional and forward follow-up study and collected the duration times between departing Wuhan and the onset of symptoms for 1211 confirmed cases in people who left Wuhan before the lockdown with no symptom of COVID-19 and then developed symptoms outside Wuhan; more details of the study and data collection can be found in Section~\ref{sec4}. By utilizing the theory of renewal processes, they proposed a \emph{mixture forward--incubation-time epidemic model} to model the 1211 observed duration times and to estimate the incubation distribution. This mixture model overcomes the issues of biased sampling and accounts for the possibility that some patients may have been exposed to COVID-19 on their way out of Wuhan.

Herein, we follow the approach and model setup of \cite{Qin2020}. Let $Y$ be the \emph{incubation period} with probability density function (pdf) $f(t)$. Consider a specific observation time that is either (i) the time of exposure to the disease or (ii) some time thereafter but before the onset of symptoms, but whether the situation pertains to (i) or (ii) is unknown. For example, \cite{Qin2020} chose the observation time of an individual to be their departure time from Wuhan. Furthermore, let $V$ be the \emph{forward time} calculated from a specific observation time to the symptom-onset time given that the observation time is after the exposure time but before the symptom-onset time. When a renewal process reaches equilibrium, the pdf of $V$ is
$$
g(t)= \frac{ {\int_{t}^\infty f(y) dy}} {\int_{0}^\infty y f(y) dy}
\quad
\mbox{for}
~ t>0
$$
\citep[Chapter 2]{Linton2020, Qin2017}.
See Section S1 of the supplement for
a derivation of the form of $g(t)$.
As \cite{Qin2020} pointed out, the study cohort may contain heterogeneous subpopulations: individuals who left Wuhan by train, bus, or plane were likely to have come into contact with COVID-19 because they were in a crowded environment with possible human-to-human transmission of the virus.
A similar argument pertains to the COVID-19 outbreak that occurred from late January to early February in 2020 onboard the Diamond Princess cruise ship \citep{Verity2020}.

In the following, we use the duration-time data from Wuhan in \cite{Qin2020} as an illustration to introduce the mixture forward--incubation-time epidemic model, in which the observation time of an individual is their departure time from Wuhan. Let $T$ be the \emph{duration time} between departure from Wuhan and the onset of symptoms. Furthermore, denote $p$ as the proportion of individuals who contracted COVID-19 as they left Wuhan. For this portion of individuals, the departure time is just the exposure time to COVID-19, and hence $T$ is the incubation period; for the other portion of individuals, the  departure time is after the exposure time to COVID-19 but before the onset of symptoms, and hence $T$ is the forward time. Because we have no idea who contracted the disease before departure and who did so while departing, $T$ follows the \emph{mixture forward--incubation-time epidemic model} \citep{Qin2020}
\begin{equation}
\label{model.general}
h(t)=p f(t)+(1-p) g(t), \quad t>0.
\end{equation}
Note that we can observe only $T$ and not $Y$ or $V$. Let $t_1,\ldots,t_n$ be $n$ observed duration times that are independent and identically distributed (iid) copies of $T$.

We point that there may exist a third portion of individuals who got infected outside Wuhan after the departure.
In this paper, we assume that this  portion of individuals does not exist for two reasons.
First, it is theoretically challenging to derive the pdf of duration time for this portion of individuals.
Some additional work is required.
The results developed under the model \eqref{model.general} can serve as a starting point  for further research.
Second, the goodness-of-fit test in Section S2 of the supplement seems to suggest that the model \eqref{model.general} provides an adequate fit to the duration-time data from Wuhan.

Throughout the paper, we focus on model \eqref{model.general} with $f(t)=f(t;\lambda,\alpha)$, the pdf of a general two-parameter distribution. Then the pdf of $T$ becomes
\begin{equation}
\label{model}
h(t;\lambda,\alpha,p)= p f(t;\lambda,\alpha)+(1-p) g(t;\lambda,\alpha),\quad t>0,
\end{equation}
and $t_1,\ldots,t_n$ are $n$ iid observations from $h(t;\lambda,\alpha,p)$. Under the mixture model \eqref{model}, \cite{Deng2020} discussed the asymptotic properties of the maximum likelihood estimators (MLEs) and the likelihood ratio statistic of unknown parameters $(\lambda,\alpha,p)$ under the assumption that $(\lambda,\alpha,p)$ are identifiable. However, this assumption does not always hold. A counter example is the Weibull pdf $f(t;\lambda,\alpha)=\lambda\alpha(t\lambda)^{\alpha-1}\exp\{-(\lambda t)^{\alpha}\} I(t>0)$. It can be verified that $f(t;\lambda,\alpha)=g(t;\lambda,\alpha)$ when $\alpha=1$. This implies that $p$ is not identifiable in \eqref{model} when $f(t;\lambda,\alpha)$ is a Weibull pdf with $\alpha=1$. Because of that, the asymptotic results in \cite{Deng2020} are not applicable in such a situation. A similar conclusion also holds when $f(t;\lambda,\alpha)= \{ {\Gamma(\alpha)} \}^{-1} {\lambda^\alpha t^{\alpha-1}\exp(-\lambda t)} I(t>0)$, a Gamma pdf.

In this paper, we complement \cite{Deng2020} in two ways. First, we provide sufficient conditions for the identifiability of $(\lambda,\alpha,p)$, and our results indicate the following: (i) $(\lambda,\alpha,p)$ is identifiable when $f(t;\lambda,\alpha)$ is a lognormal pdf, and when $f(t;\lambda,\alpha)$ is a Weibull or Gamma pdf but not an exponential pdf; (ii) $(\lambda,\alpha)$ is identifiable but $p$ is not when $f(t;\lambda,\alpha)$ is an exponential pdf. Second, we propose a likelihood ratio test (LRT) to test the null hypothesis that $f(t;\lambda,\alpha)$ is an exponential pdf.
Under this null hypothesis, $h(t;\lambda,\alpha,p)$ also becomes an exponential pdf, so the proposed LRT also tests the homogeneity in model \eqref{model}.
Note that the nuisance parameter $p$ disappears under the null model and is only identified under the alternative hypothesis.

The problem of a nuisance parameter unidentified under the null hypothesis has long been recognized in the literature as a non-regular problem \citep{Davies1977,Davies1987}. Because of the partial identifiability of the nuisance parameter, classical inference methods such as the LRT may lose their usual statistical properties. The limiting distribution of the LRT often involves complex stochastic processes \citep{Liu2020-SJS}. The homogeneity testing problem under a two-component mixture model has been studied extensively in the literature; for example, see  \cite{Liu2003}, \cite{Chen2009}, and \cite{Chen2020} and the references therein. To the best of our knowledge, these papers assume that the two components come from the same distribution family and do not share any underlying parameters. However, under model \eqref{model}, the two components are not from the same distribution family and  share the common parameters $(\lambda,\alpha)$, so the existing results cannot be applied to the testing problem under model \eqref{model}.

Despite the aforementioned challenges, we successfully work out the limiting distribution of the LRT for the non-regular testing problem, i.e., testing the null hypothesis that $h(t;\lambda,\alpha,p)$ is the pdf of a homogeneous exponential distribution. We show that the asymptotic null distribution of the LRT is the supremum of a chi-square process, and further we identify an explicit representation of the limiting distribution that can be used for rapid numerical calculation of the asymptotic critical values or $p$-values of the proposed LRT. By extensive simulations, we find that the proposed LRT has desirable finite-sample testing performance, i.e., tight control of type I error rates and appreciable powers in general. The proposed LRT  is then used to analyze COVID-19 data from China for illustration. Following \cite{Qin2020}, we choose $f(t;\lambda,\alpha)$ to be a Weibull pdf, and the analysis results indicate that the mixture forward--incubation-time model produces a better fit than that with a homogeneous exponential distribution.

Note that all the results herein are based on parametric model \eqref{model}, and violation of this model assumption may lead to invalid subsequent analysis results. This raises the goodness-of-fit test problem of model \eqref{model} in applications. We suggest using the goodness-of-fit test in \cite{Deng2020} to check the validity of model \eqref{model} based on $t_1,\ldots,t_n$, and this test is reviewed briefly in the supplement for presentational completeness.

The rest of this paper is organized as follows. In Section~\ref{sec2.1}, we discuss sufficient conditions for the identifiability of $(\lambda,\alpha,p)$ in model \eqref{model}, and we apply the results to the case where $f(t;\lambda,\alpha)$ is a Weibull, Gamma, or lognormal pdf. In Section~\ref{sec2.2}, we establish the non-regular asymptotic distribution of the LRT for testing the null hypothesis that $h(t;\lambda,\alpha,p)$ is a homogeneous exponential distribution and we also provide an explicit representation of this asymptotic distribution. The asymptotic distribution of the proposed LRT under a sequence of local alternatives is also derived. We report our simulation results in Section~\ref{sec3}, and in Section \ref{sec4} we analyze real COVID-19 outbreak data from China for illustration. Finally, we conclude the paper with a discussion in Section \ref{sec5}. For convenience of presentation, all proofs are given in the supplementary material.

\setcounter{section}{1} 
\setcounter{equation}{0} 

\section{Identifiability of $(\lambda,\alpha,p)$\label{sec2.1}}
Identifiability is an important issue in the application of the mixture forward--incubation-time epidemic model in \eqref{model}. If some model parameters are not identifiable, then their point estimators cannot be consistent, and standard inferences for other parameters that are identifiable may be questionable. In this section, we establish the identifiability of $(\lambda,\alpha,p)$ in model \eqref{model} under the following conditions on $f(t;\lambda,\alpha)$.
Let $F(t;\lambda,\alpha)$ be the cumulative distribution function corresponding to $f(t;\lambda,\alpha)$.
\begin{enumerate}
\item[A1.]
Given $(\lambda,\alpha)$, $\lim_{t\to \infty}\frac{f(t;\lambda,\alpha)}{1-F(t;\lambda,\alpha)}$ exists and is either finite or $\infty$.
\item[A2.]
When $(\lambda_1,\alpha_1)\neq (\lambda_2,\alpha_2)$, $\lim_{t\to \infty}\frac{f(t;\lambda_1,\alpha_1)}{f(t;\lambda_2,\alpha_2)}$ exists and is either 0 or $\infty$.
\item[A3.]
When $(\lambda_1,\alpha_1)\neq (\lambda_2,\alpha_2)$, both $\lim_{t\to \infty}\frac{f(t;\lambda_1,\alpha_1)}{1-F(t;\lambda_2,\alpha_2)}$ and $\lim_{t\to \infty}\frac{f(t;\lambda_2,\alpha_2)}{1-F(t;\lambda_1,\alpha_1)}$ exist and are either 0 or $\infty$. \end{enumerate}

\begin{theorem}
\label{prop1}
Assume model \eqref{model} and conditions A1--A3. Let
$$
A(\lambda,\alpha)=\lim_{t\to \infty}\frac{f(t;\lambda,\alpha)}{1-F(t;\lambda,\alpha)}.
$$
Suppose $h(t;\lambda_1,\alpha_1,p_1)=h(t;\lambda_2,\alpha_2,p_2)$ for all $t>0$.
\begin{itemize}
\item[(a)] If $A(\lambda_1,\alpha_1)=0$ or $\infty$, then $(\lambda_1,\alpha_1,p_1)=(\lambda_2,\alpha_2,p_2)$.
\item[(b)]
If $0<A(\lambda_1,\alpha_1)<\infty$, then $(\lambda_1,\alpha_1)=(\lambda_2,\alpha_2)$. Furthermore, if $\frac{f(t;\lambda_1,\alpha_1)}{1-F(t;\lambda_1,\alpha_1)}$ is not a constant function of $t$, then $p_1=p_2$; otherwise, $p_1$ and $p_2$ are not necessarily the same.
\end{itemize}
\end{theorem}

After some calculus work, it can be verified that conditions~A1--A3 are satisfied by a Weibull, Gamma, or lognormal distribution. We can further verify that $A(\lambda,\alpha)=0$ for a lognormal distribution, $A(\lambda,\alpha)=\lambda$ for a Gamma distribution, and $A(\lambda,\alpha)=0$ or $\infty$ if $\alpha\neq 1$ and $A(\lambda,\alpha)=\lambda$ if $\alpha=1$ for a Weibull distribution. Applying the results in Theorem~\ref{prop1} to Weibull, Gamma, and lognormal distributions, we have the following identifiability results.

\begin{corollary}
\label{coro1}
Under model \eqref{model},
\begin{itemize}
\item[(a)] $(p,\lambda,\alpha)$ are identifiable when $f(t;\lambda,\alpha)$ is the pdf of a lognormal distribution;
\item[(b)] $(p,\lambda,\alpha)$ are identifiable when $f(t;\lambda,\alpha)$ is the pdf of a Weibull or Gamma distribution but not the pdf of an exponential distribution;
\item [(c)] $(\lambda,\alpha)$ are identifiable but $p$ is not when $f(t;\lambda,\alpha)$ is the pdf of an exponential distribution.
\end{itemize}
\end{corollary}

\cite{Deng2020} mentioned the identifiability property of $(\lambda,\alpha,p)$ but did not {give a formal proof.} The results in Theorem~\ref{prop1} and Corollary~\ref{coro1} provide formal justifications and further indicate when the results of \cite{Deng2020} are applicable and when they are not.

\section{Testing Whether Incubation Distribution is Exponential \label{sec2.2}}
\subsection{Likelihood Ratio Test}
Corollary~\ref{coro1} indicates that the parameter $p$ is not identifiable when $f(t;\lambda,\alpha)$ is the pdf of an exponential distribution under model \eqref{model}. Because of this, the asymptotic results in \cite{Deng2020} are not applicable in such a situation. In this section, we propose an LRT to check whether $f(t;\lambda,\alpha)$ is the pdf of an exponential distribution, or equivalently whether $h(t;\lambda,\alpha,p)$ is the pdf of a homogeneous exponential distribution, based on $n$ iid observations $t_1,\ldots,t_n$ from model \eqref{model}.

Throughout this section, we assume that the following condition is satisfied.
\begin{itemize}
\item[C0.]There exists a unique $\alpha_0$ such that $f(t;\lambda,\alpha_0)=g(t;\lambda,\alpha_0)$ for all $t>0$.
\end{itemize}
Condition~C0 is satisfied by a Weibull or Gamma distribution with $\alpha_0=1$ in each case, and it can be shown that condition~C0 is satisfied if and only if $f(t;\lambda,\alpha_0)$ is the pdf of an exponential distribution. Under condition~C0, testing the null hypothesis that $f(t;\lambda,\alpha)$ is the pdf of an exponential distribution is equivalent to testing
\ba
\label{H0}
H_{0}: \alpha = \alpha_0\quad \mbox{versus} \quad H_{1}: \alpha\neq \alpha_0.
\ea
Note that under model \eqref{model}, the case of $\alpha = \alpha_0$ indicates that individuals in the cross-sectional and forward follow-up study are homogeneous, and the duration time $T$ defined in Section~\ref{sec1} follows an exponential distribution. When $\alpha\neq \alpha_0$, there are heterogeneous subgroups of individuals in the cross-sectional and forward follow-up study. In this case, we favor using the mixture model \eqref{model} to model the distribution of $T$. Theoretically, detecting the existence of such heterogeneous subpopulations is an important initial step before applying the mixture model \eqref{model}. If we were to apply model \eqref{model} to homogenous duration times, then the MLE of $(\lambda,\alpha,p)$ would no longer have asymptotic normality, and consequently the Wald-type confidence intervals for the quantiles of the incubation period may not have the nominal asymptotic coverage probabilities.

A natural solution to the testing problem \eqref{H0} is one based on likelihood. Given the $n$ observations $t_1,\ldots,t_n$ from model \eqref{model}, the log-likelihood of $ (\lambda, \alpha, p)$ is
\bas
\ell_n(\lambda, \alpha, p)
&=& \sum_{i=1}^n \log \left\{ p f(t_i;\lambda,\alpha)+(1-p) g(t_i;\lambda,\alpha) \right\}.
\eas
Let $(\hat\lambda,\hat\alpha,\hat p)$ be the MLE of $(\lambda, \alpha, p)$ under the full model, and let $\hat\lambda_0$ be the MLE of $\lambda$ under the null model, {i.e.,}
$$
(\hat\lambda,\hat\alpha,\hat p)
=\arg\max_{\lambda,\alpha,p} \ell_n(\lambda, \alpha, p),
\quad
\hat\lambda_0=\arg\max_{\lambda} \ell_n(\lambda, \alpha_0, 1).
$$
{Note that under the null model, $p$ does not appear and $\lambda$ is the only parameter to be estimated.} We simply set $p=1$ under the null model for convenience of presentation.

The LRT statistic for \eqref{H0} is defined as
$$
R_n=2\left\{\sup_{\lambda, \alpha, p}\ell_n(\lambda, \alpha, p)-
\sup_{\lambda}\ell_n(\lambda, \alpha_0, 1)
\right\}=2\left\{ \ell_n(\hat\lambda,\hat\alpha,\hat p)-\ell_n(\hat\lambda_0,\alpha_0,1)\right\}.
$$
We reject the null hypothesis $H_0$ in \eqref{H0} if the observed value of $R_n$ exceeds some critical value determined by its limiting distribution presented in Section~\ref{sec3.2}.

\subsection{Asymptotic Null Distribution of Likelihood Ratio Test \label{sec3.2}}
We require some notation before presenting the asymptotic results of the LRT statistic $R_n$. Let $(\lambda_0,\alpha_0)$ be the true value of $(\lambda,\alpha)$ under the null model, and define
\begin{eqnarray*}
X_i=\frac{\partial f(t_i;\lambda_0,\alpha_0)/\partial \lambda}{f(t_i; \lambda_0,\alpha_0)},~~
Y_{i1}=\frac{\partial f(t_i;\lambda_0,\alpha_0)/\partial \alpha}{f(t_i; \lambda_0,\alpha_0)},~~
\label{Yi1}
Y_{i2}=\frac{\partial g(t_i;\lambda_0,\alpha_0)/\partial \alpha}{g(t_i; \lambda_0,\alpha_0)}.
\label{Yi2}
\end{eqnarray*}
Note that under condition~C0,
$$
f(t_i;\lambda_0,\alpha_0)=g(t_i;\lambda_0,\alpha_0)~~
\mbox{ and }
~~
\frac{\partial g(t_i;\lambda_0,\alpha_0)/\partial \lambda}{g(t_i; \lambda_0,\alpha_0)}=X_i.
$$
Define ${\bf b}_i=(X_i,Y_{i1},Y_{i2})^\T$ and denote the variance-covariance matrix
\begin{equation}
{\bf B}=\var({\bf b}_i)
=\left(
\begin{array}{ccc}
B_{11}&B_{12}&B_{13}\\
B_{21}&B_{22}&B_{23}\\
B_{31}&B_{32}&B_{33}\\
\end{array}
\right),
\end{equation}
where the variance is taken with respect to the null model. Furthermore, define
\begin{eqnarray*}
\sigma_{11}&=&B_{33}-\frac{B_{13}^2}{B_{11}},\quad
\sigma_{12}~=~B_{23}-B_{33}-\frac{B_{12}B_{13}}{B_{11}}+\frac{B_{13}^2}{B_{11}},\\
\sigma_{22}&=&B_{22}+B_{33}-2B_{23}-\frac{B_{12}^2}{B_{11}}-\frac{B_{13}^2}{B_{11}}+ \frac{2B_{12}B_{13}}{B_{11}}.
\end{eqnarray*}
For any $p_1,p_2\in[0,1]$, let
\begin{equation}
\label{general.sigma}
\sigma(p_1,p_2)=p_1p_2\sigma_{22}+(p_1+p_2)\sigma_{12}+\sigma_{11}.
\end{equation}

Our asymptotic results about $R_n$ rely on conditions C1--C5 given in Section S3 of the supplement;
they are typical regularity conditions in the literature of finite mixture models.

\begin{theorem}
\label{thm1}
Suppose that conditions~C0 and C1--C5 in the supplement are satisfied. Under model \eqref{model} and the null hypothesis in \eqref{H0}, as $n\to\infty$,
\bas
R_n \rightarrow R=\sup\limits_{0\leq p \leq 1} Z^2(p)
\eas
in distribution, where $Z(p)$ is a Gaussian process with zero mean, unit variance, and covariance function
\begin{equation*}
\cov\big\{Z(p_1), Z(p_2)\big\}
=\frac{\sigma(p_1,p_2)}{\sqrt{{\sigma(p_1,p_1) \sigma(p_2,p_2) }}}, \quad 0\leq p_1, p_2\leq 1.
\end{equation*}
\end{theorem}

Theorem~\ref{thm1} shows that the LRT statistic $R_n$ has a non-regular limiting distribution that is the supremum of a $\chi^2$-process. In general, the distribution function of $R$, i.e., the supremum of a $\chi^2$-process, does not have a closed form and is difficult to calculate numerically. Instead, we derive an equivalent representation of $R$ that is much simpler in form, and with which it is much more convenient to calculate the distribution function or quantiles of $R$ by the Monte~Carlo method.

\begin{figure}[!htt]
\vspace{-0.3in}
\centerline{\epsfig{file=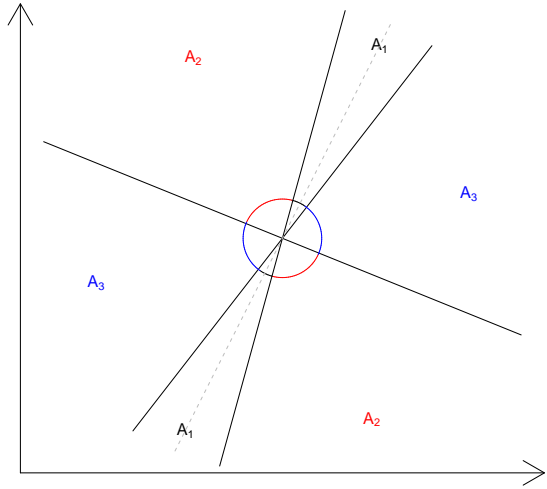,width=3.6in}}
\vspace{-0.7in}
\caption{Graphical representation of sets $A_1$, $A_2$, and $A_3$ when $f(t;\lambda,\alpha)$ is a Weibull pdf.}
\label{fig1}
\end{figure}

We require some additional notation. Consider the following polar transformation:
$
(\cos\theta, \sin\theta)=\big(c_1(p),c_2(p)\big),
$
where
$$
c_1(p)=\frac{\sqrt{\sigma_{11}-\sigma_{12}^2/\sigma_{22}}}{\sqrt{\sigma(p,p)}}
~~
\mbox{ and }
~~
c_2(p)=\frac{(p+\sigma_{12}/\sigma_{22})\sqrt{\sigma_{22}}}{\sqrt{\sigma(p,p)}}.
$$
To find a simple representation for $R$, we require the following additional condition.
\begin{enumerate}
\item[C6.] There exist $\Delta_1$ and $\Delta_2$
such that $-\pi/2< \Delta_1<\Delta_2<\pi/2$ and
$$
\left\{\big(c_1(p),c_2(p)\big):0\leq p \leq 1\right\}=\{(\cos\theta,\sin\theta):
\Delta_1\leq\theta\leq\Delta_2\}.
$$
\end{enumerate}
Under condition~C6, we define the three sets
\begin{align*}
&A_1=\{\eta: \max_{\theta\in[\Delta_1,\Delta_2]} \cos^2(\theta-\eta)=1 \}, \nonumber \\
&A_2=\{\eta: \max_{\theta\in[\Delta_1,\Delta_2]} \cos^2(\theta-\eta)= \cos^2(\eta-\Delta_2) \}
, \nonumber \\
&A_3=\{\eta: \max_{\theta\in[\Delta_1,\Delta_2]} \cos^2(\theta-\eta)= \cos^2(\eta-\Delta_1) \}.
\end{align*}
If both $\Delta_1$ and $\Delta_2$ are positive, then these sets have the following explicit forms:
\begin{align}
&A_1=[\Delta_1, \Delta_2]
\cup[\Delta_1-\pi, \Delta_2-\pi], \nonumber \\
&A_2=[\Delta_2,\Delta+\pi/2]
\cup[\Delta_2-\pi,\Delta-\pi/2], \nonumber \\
&A_3=[\Delta+\pi/2,\pi]
\cup[-\pi,\Delta_1-\pi]\cup[\Delta-\pi/2,\Delta_1],
\label{set-A}
\end{align}
where $\Delta=(\Delta_1+\Delta_2)/2$. Figure~\ref{fig1} shows $A_1$--$A_3$ graphically when $f(t;\lambda,\alpha)$ is a Weibull pdf.

\begin{theorem}
\label{thm2}
Assume the conditions of Theorem~\ref{thm1} and condition~C6. Furthermore, suppose that $\rho^2$ and $\eta$ are two independent random variables that follow $\chi_2^2$ and the uniform distribution on $[-\pi,\pi]$, respectively. Then $R$ has the same distribution as
\begin{equation*}
\label{rep2}
T(\rho^2, \eta) =
\rho^2 \{ I (\eta\in A_1) + I (\eta\in A_2) \cos^2(\eta-\Delta_2) + I (\eta\in A_3) \cos^2(\eta-\Delta_1) \}.
\end{equation*}
\end{theorem}

Note that $\Delta_1$, $\Delta_2$, and $A_1$--$A_3$ may depend on $\lambda_0$. We can estimate $\lambda_0$ using $\hat\lambda_0$, the MLE of $\lambda$ under the null model. Based on Theorem~\ref{thm2}, we propose the following Monte~Carlo procedure for approximating the distribution and quantiles of $R$. First, we generate a large number (e.g., $M=10^8$) of independent copies of $(\rho^2, \eta)$, denoted by $(\rho_i^2, \eta_i)$ ($i=1, \ldots, M$). Then, we take the empirical distribution of $\{T(\rho_i^2, \eta_i), i=1, \ldots, M \}$ to approximate the distribution of $R$. Accordingly, we can calculate the approximate $p$-value of the LRT or the approximate quantiles of $R$, which may serve as critical values of the proposed LRT.

The results in Theorems~\ref{thm1} and \ref{thm2} rely on the forms of $\sigma(\cdot,\cdot)$ in \eqref{general.sigma} and $(\Delta_1,\Delta_2)$ in condition~C6. In the following, we identify two examples satisfying conditions~C0--C6 and work out their $\sigma(\cdot,\cdot)$ and $(\Delta_1,\Delta_2)$.

\begin{example} (Weibull distribution). Recall that the pdf of a Weibull distribution is given as $f(t;\lambda,\alpha)=\lambda\alpha(t\lambda)^{\alpha-1}\exp\{-(\lambda t)^{\alpha}\} I(t>0)$. It can be shown that
$
\sigma({p_1,p_2})
=p_1p_2(\pi^2/6-1)+ (p_1+p_2)(2-\pi^2/6)+ \pi^2/3-3
$
and
$$
\Delta_1=\arccos\left(\sqrt{{\frac{\pi^4-6\pi^2-36}{2\pi^4-30\pi^2+108}}}\right),
\quad
\Delta_2=\arccos\left(\sqrt{{\frac{\pi^4-6\pi^2-36}{\pi^4-6\pi^2}}}\right).
$$
Because both $\Delta_1$ and $\Delta_2$ are positive, $A_1$--$A_3$ take the forms in \eqref{set-A}.
\end{example}

\begin{example} (Gamma distribution). Recall that the pdf of a Gamma distribution is given as $f(t;\lambda,\alpha)= \{\Gamma(\alpha)\}^{-1}\lambda^\alpha t^{\alpha-1}\exp(-\lambda t) I(t>0)$. It can be shown that
$
\sigma(p_1,p_2)=p_1p_2\left(\frac{\pi^2}{6}-\frac{5}{4}\right)
+(p_1+p_2) \left(\frac{7}{4}-\frac{\pi^2}{6}\right)+\frac{\pi^2}{3}-\frac{13}{4}
$
and
\begin{eqnarray*}
\Delta_1
&=&\arccos\left(\sqrt{{\frac{4\pi^4-54\pi^2+144}{(4\pi^2-39)(2\pi^2-15) }}}\right),
\\
\Delta_2
&=&\arccos\left(\sqrt{{\frac{4\pi^4-54\pi^2+144}{(2\pi^2-12)(2\pi^2-15)}}}\right).
\end{eqnarray*}
Again, both $\Delta_1$ and $\Delta_2$ are positive, so $A_1$--$A_3$ again take the forms in \eqref{set-A}.
\end{example}

As we can see, $\sigma(\cdot,\cdot)$ and $(\Delta_1,\Delta_2)$ for a Weibull or Gamma distribution are independent of $\lambda_0$, so there is no need to estimate $\lambda_0$ when using Theorem~\ref{thm2} for these two distributions.

\subsection{Asymptotic Power of Likelihood Ratio Test}
In this subsection, we study the asymptotic power of the proposed LRT. We consider the following sequence of local alternatives that are indexed by $n$:
\begin{equation}
\label{local.alt}
H_a^n: \lambda=\lambda_0, ~p=p_0, ~\alpha=\alpha_0+\delta n^{-1/2},
\end{equation}
where $\delta$ is a fixed constant independent of $n$. The following theorem presents the asymptotic distribution of $R_n$ under $H_a^n$.

\begin{theorem}
\label{thm3}
Assume the conditions of Theorem~\ref{thm1}. Under the local alternative hypothesis $H_a^n$ in \eqref{local.alt}, as $n\to\infty$,
\begin{equation}
\label{local.power}
R_n \to \sup\limits_{0\leq p \leq 1}\left[ \left\{Z(p)+\omega(p,p_0)\right\}^2\right]
\end{equation}
in distribution, where $\omega(p,p_0)=\delta\sigma ({p,p_0})/\sqrt{\sigma ({p,p}) }$ and $Z(p)$ is defined in Theorem \ref{thm1}.
\end{theorem}

Note that the result in Theorem~\ref{thm3} has two important applications. First, it is useful for local power analysis for a potential alternative model with the model parameters $(\lambda,\alpha,p)$. We can insert this model into the local sequence and obtain $\delta=n^{1/2}(\alpha-\alpha_0)$, and the power of $R_n$ for detecting this alternative model can then be assessed based on the limiting distribution under the local alternative. Second, the result in Theorem~\ref{thm3} also sheds light on the power trend under different alternative models; for example, if $f(t;\lambda,\alpha)$ is the pdf of a Weibull distribution, then $|\omega(p,p_0)|$ increases as $\delta$ departs from zero or $p_0$ increases. This implies that the power of $R_n$ increases as $\alpha$ departs from $\alpha_0=1$ and/or the value of $p$ under the alternative model increases. This trend is confirmed in the following simulation study.

\section{Simulation}
\label{sec3}
\begin{figure}[!htt]
\centerline{\epsfig{file=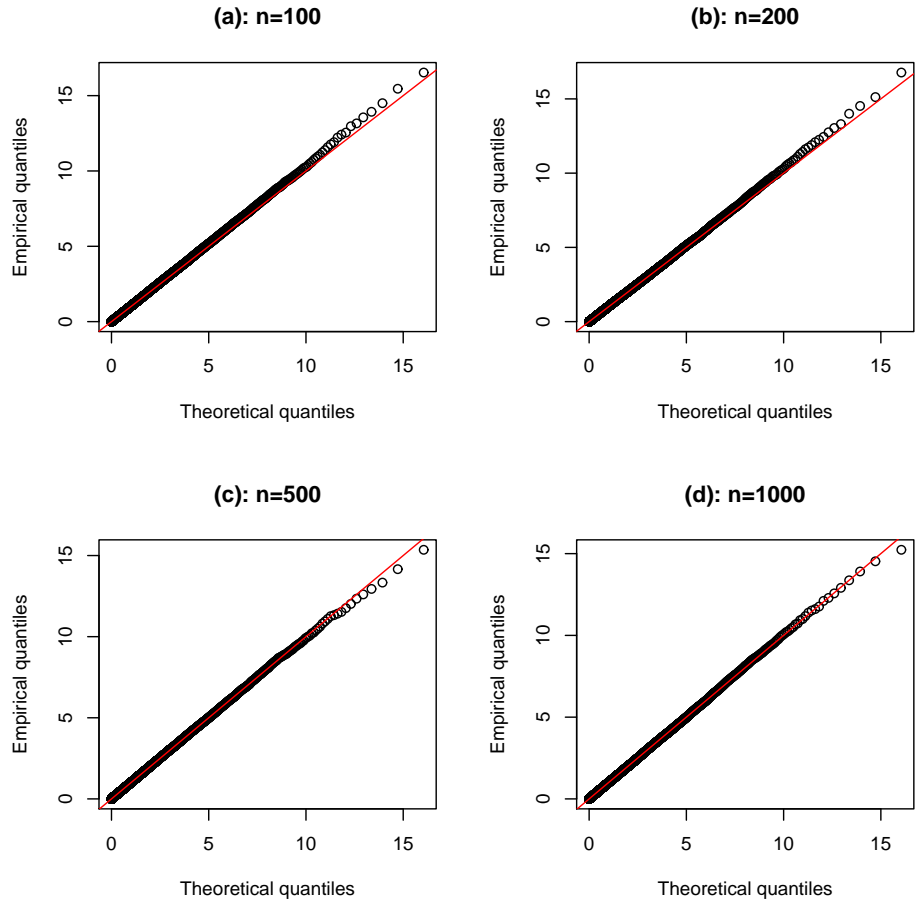,width=5in}}\par
\caption{Quantile-quantile plots of $R_n$ {for different sample sizes.}}
\label{fig2}
\end{figure}

In this section, we use simulations to check whether the limiting distribution of $R_n$ provides an accurate approximation to its finite-sample distribution. We consider four sample sizes: $n=100$, 200, 500, and 1000. Following \cite{Qin2020}, we choose $f(t;\lambda,\alpha)$ to be a Weibull pdf and set the true value of $\lambda$ to be 1. Note that under $H_0$ in \eqref{H0}, the true value of $\alpha$ is 1 and $p$ disappears. The simulated type I errors of $R_n$ based on $10^5$ repetitions are summarized in Table~\ref{typeI}. The simulation results show that the proposed LRT test has tight control of type I error rates for all the combinations of sample size and significance level. Figure~\ref{fig2} shows the quantile-quantile plots of the LRT test. As can be seen, the limiting null distribution of $R_n$ provides an adequate approximation to its finite-sample distribution even when the sample size is as small as 100.

\begin{table}[!ht]
\caption{
Type I error rates (in \%) of $R_n$ at a significance level of 10\%, 5\%, or 1\%.
\label{typeI}}
\centering
\tabcolsep 19pt
\renewcommand{\arraystretch}{0.75}
\begin{tabular}{ c ccc }
\hline
$n$ & \multicolumn{3}{c }{Significance level} \\
& $10\%$ & $5\%$ & $1\%$ \\ \hline
100 &10.6&5.4&1.1\\
200 &10.2&5.2&1.1\\
500 &10.1&5.1&1.0\\
1000&10.1&5.0&1.0\\
\hline
\end{tabular}
\end{table}

Next, we evaluate the power of the proposed LRT test. We consider two true values of $\alpha$ equal to 1.35 and 1.65, and three true values of $p$ equal to 0.15, 0.40 and 0.65, and the simulated powers based on $10^4$ repetitions are summarized in Table~\ref{power}. We observe that the proposed LRT test has appreciable powers in all the cases considered. Furthermore, its power increases as $p$ or $\alpha$ increases, and this trend agrees with the local power analysis after Theorem~\ref{thm3}.

\begin{table}[!ht]
\caption{
Power (in \%) of $R_n$ at a significance level of 10\%, 5\%, or 1\%.
\label{power}}
\centering
\tabcolsep 15pt
\renewcommand{\arraystretch}{0.75}
\begin{tabular}{ ccccccc }
\hline
$n$ & \multicolumn{3}{c }{Significance level} & \multicolumn{3}{c }{Significance level} \\
& $10\%$ & $5\%$ & $1\%$ & $10\%$ & $5\%$ & $1\%$ \\
\hline
& \multicolumn{3}{c }{$(p,\alpha)=(0.15,1.35)$} & \multicolumn{3}{c }{$(p,\alpha)=(0.15,1.65)$} \\
100 &~58.4&~45.3&22.4&~89.7&~81.9&~59.5\\
200 &~81.9&~72.2&47.5&~99.2&~98.2&~92.5\\
500 &~99.2&~98.1&92.0&100.0&100.0&100.0\\
1000&100.0&100.0&99.9&100.0&100.0&100.0\\
\hline
& \multicolumn{3}{c }{$(p,\alpha)=(0.40,1.35)$} & \multicolumn{3}{c }{$(p,\alpha)=(0.40,1.65)$} \\
100 &~76.7&~65.3&~39.4&~97.8&~95.4&~84.7\\
200 &~95.0&~90.4&~74.0&100.0&100.0&~99.5\\
500 &100.0&~99.9&~99.6&100.0&100.0&100.0\\
1000&100.0&100.0&100.0&100.0&100.0&100.0\\
\hline
& \multicolumn{3}{c }{$(p,\alpha)=(0.65,1.35)$} & \multicolumn{3}{c }{$(p,\alpha)=(0.65,1.65)$} \\
100 &~90.2&~82.7&~60.2&~99.9&~99.7&~97.9\\
200 &~99.4&~98.6&~93.0&100.0&100.0&100.0\\
500 &100.0&100.0&100.0&100.0&100.0&100.0\\
1000&100.0&100.0&100.0&100.0&100.0&100.0\\
\hline
\end{tabular}
\end{table}

\section{Application to COVID-19 Data}
\label{sec4}
The outbreak of COVID-19 in Wuhan, China in December 2019 attracted worldwide attention \citep{Li2020,Wang2020,Tu2020}. To prevent {its spread} before being out of control, the Chinese government decided to {lock down} Wuhan on January 23, 2020. From public reports, there were many confirmed cases of people who left Wuhan before the lockdown with no symptoms of COVID-19 but who then developed symptoms outside Wuhan.

\cite{Deng2020} provided data based on confirmed cases of COVID-19 reported in publicly available sources such as provincial and municipal health commissions in China and the health authorities in other countries as of February 15, 2020. The duration time for a patient was recorded as the time difference between leaving Wuhan and the earliest onset of symptoms (e.g., fever, cough). Our analysis involves a sample size of 1211 cases and satisfies the design criteria of the mixture forward--incubation-time epidemic model \eqref{model}. These criteria include the following. (1) The included cases were of people who had no COVID-19 symptoms when they left Wuhan and developed symptoms elsewhere after traveling. Hence, cases of people whose first symptoms occurred before traveling were not included in the sample. (2) The date of leaving Wuhan had to be between January 19, 2020 and January 23, 2020 for the following reasons: (2a) before January 19, 2020, the public were as yet unaware of the severity of COVID-19, so there may have been a chance that a patient was actually infected outside Wuhan after they left; (2b) after January 23, 2020 (the date of the Wuhan lockdown), there were not many cases available, and also this enabled us to have an average follow-up time for symptoms onset of as long as 25 days. This sample size of 1211 is relatively large compared with other incubation period studies of COVID-19.

Following \cite{Qin2020}, we use model \eqref{model} with $f(t;\lambda,\alpha)$ being a Weibull pdf to analyze the 1211 observed duration times. At the beginning of the outbreak, it was more likely to observe someone who had been infected closer to their departure date as the number of infections grew exponentially, and this may invalidate the assumptions for deriving the forward time distribution \citep{Qin2020, Liu2020AMAS}. Because of that, we may be concerned about the validity of the model assumptions in \eqref{model} for the 1211 observed duration times. To address this concern, \cite{Deng2020} performed a goodness-of-fit test for model \eqref{model}. The asymptotic $p$-value of this test is found to be 0.37, which indicates that model \eqref{model} with $f(t;\lambda,\alpha)$ being a Weibull pdf provides a reasonable fit to the 1211 observed duration times; see the supplement for more details. Next, we test for $\alpha=1$, or equivalently, whether the data come from a homogeneous exponential distribution, by using the proposed LRT when $f(t;\lambda,\alpha)$ is a Weibull pdf.

All the observed duration times are integers of between zero and 22 days, and in theory our proposed method may not be directly applicable. For illustration, we impute the value of observed integer value $i$ by a random number from ${\rm U}(i,i+1)$, the uniform distribution on $(i,i+1)$; for example, the frequency for zero days is 82, so we generate 82 observations from ${\rm U}(0,1)$. After that, we apply the proposed testing procedure to the imputed data set. We repeat the procedure 1000 times and obtain 1000 estimates of $(\lambda,\alpha,p)$ and 1000 LRT statistics $R_n$. Based on these 1000 repetitions, the average of the estimates for $(\lambda,\alpha,p)$ is $(0.655,0.135,1.645)$. The values of $R_n$ range from 202.9 to 234.3, and because the $p$-value of any LRT statistic in $[202.9,234.3]$ is almost zero, this provides overwhelming evidence for rejecting the null hypothesis of $\alpha=1$.

We have also analyzed the data after adding 0.5 to each duration time, i.e., any integer datum $i$ is replaced with the mid-point of the interval $(i,i+1)$. The resulting $R_n$ is around 230.7, with a $p$-value still of almost zero. From both analyses, we conclude from highly significant evidence that the population distribution of the observed duration times cannot be modeled well enough by an exponential distribution.

The above analysis results indicate that the data contain heterogeneous subgroups. Unfortunately, we have no idea who in the cohort contracted the disease before and who did so immediately upon departure, so it is more reasonable to use the mixture forward--incubation-time epidemic model \eqref{model} than a homogeneous exponential distribution to model the observed duration times.

\section{Discussion}
\label{sec5}
In this paper, we have provided sufficient conditions for the identifiability of the parameters in model \eqref{model} and applied the results to Weibull, Gamma, and lognormal distributions. We also proposed an LRT for testing the null hypothesis that $h(t;\lambda,\alpha,p)$ in \eqref{model} is the pdf of a homogeneous exponential distribution, and we derived the limiting distribution of the LRT under the null model and under a sequence of local alternatives. Our simulation results and an analysis of COVID-19 outbreak data have demonstrated the usefulness of the LRT. These results strengthen the epidemiological application of the mixture forward--incubation-time epidemic model and enrich the literature for COVID-19 data analysis.

The proposed method relies on the model assumptions in \eqref{model}. When analyzing different data sets for COVID-19 or for a new infectious virus, a goodness-of-fit test for the model assumptions in \eqref{model} is required before using the proposed LRT. We may also model the incubation-period distribution $f(t)$ nonparametrically in \eqref{model.general}. However, $(p,f)$ may not be identifiable under this setup. Some reasonable assumptions are required to ensure model identifiability, and we leave this as a future research topic.

\vskip 14pt
\noindent {\large\bf Supplementary Material}

\noindent The online supplementary material contains a derivation of the form of $g(t)$,
a goodness-of-fit test of model \eqref{model}, conditions C1--C5, and   proofs of Theorems~\ref{prop1}--\ref{thm3}.
\par


\markboth{\hfill{\footnotesize\rm Chunlin Wang, Pengfei Li, Yukun Liu, Xiao-Hua Zhou, Jing Qin} \hfill}
{\hfill {\footnotesize\rm Mixture forward--incubation-time epidemic model} \hfill}

\bibhang=1.7pc
\bibsep=2pt
\fontsize{9}{14pt plus.8pt minus .6pt}\selectfont
\renewcommand\bibname{\large \bf References}
\expandafter\ifx\csname
natexlab\endcsname\relax\def\natexlab#1{#1}\fi
\expandafter\ifx\csname url\endcsname\relax
  \def\url#1{\texttt{#1}}\fi
\expandafter\ifx\csname urlprefix\endcsname\relax\def\urlprefix{URL}\fi

 \bibliographystyle{apalike}
\bibliography{nCoV}

\vskip .65cm
\noindent
Department of Statistics and Data Science, School of Economics, Wang Yanan Institute for Studies in Economics, MOE Key Lab of Econometrics and Fujian Key Lab of Statistics, Xiamen University, Xiamen, China
\vskip 2pt
\noindent
E-mail: wangc@xmu.edu.cn
\vskip 12pt

\noindent
Department of Statistics and Actuarial Sciences, University of Waterloo, Canada
\vskip 2pt
\noindent
E-mail: pengfei.li@uwaterloo.ca
\vskip 12pt

\noindent
Key Laboratory of Advanced Theory and Application in Statistics and Data Science - MOE, and
School of Statistics, East China Normal University, Shanghai, China
\vskip 2pt
\noindent
E-mail: ykliu@sfs.ecnu.edu.cn
\vskip 12pt

\noindent
Department of Biostatistics, School of Public Health, Peking University, Beijing, China
\vskip 2pt
\noindent
E-mail: azhou@math.pku.edu.cn
\vskip 12pt

\noindent
National Institute of Allergy and Infectious Diseases, National Institutes of Health, USA
\vskip 2pt
\noindent
E-mail: jingqin@niaid.nih.gov

\newpage
\appendix

\section{Derivation of the form of $g(t)$ \label{S1}}

To write the pdf of $V$ in the form of $g(t)$, we may refer to Chapter 2 of \cite{Qin2017} using renewal process results.
Here, we may understand $V$ in the way of \cite{Linton2020}.

Let $A$ be the time elapse between exposure to the disease and the departure from Wuhan.
Recall that we use $Y$ for the incubation time, i.e., from infection onset to symptom onset.
We assume $A$ and $Y$ are independent.

By the criteria in data collection, only those individuals with $Y>A$ are included in our cohort.
Moreover, we do not observe $A$, but we can only observe $V=Y-A$.
Basically, we have a truncated data $(A,V)|Y>A$ 
Hence, the probability density function (pdf) of $V$ should be conditional on $Y>A$.

Consider the conditional cumulative distribution function (cdf) of $V$ given $Y>A$:
$$
P(V\leq t |Y>A)=\frac{P(V\leq t,Y>A)}{P(Y>A)}.
$$
%
We assume that $A$ follows uniform distribution ${\rm U}[0, c]$ for some positive constant $c$.
We further assume that $Y$  has the same support as $A$, and has the cdf and pdf, $F(t)$ and $f(t)$, respectively.
Then, conditional on $A$,
we have
\begin{eqnarray*}
P(V\leq t,Y>A)&=&\int_0^c \frac{1}{c}P(V\leq t, Y>a|A=a)da\\
&=&\int_0^c \frac{1}{c}P(Y-A\leq t, Y>a|A=a)da\\
&=&\int_0^c \frac{1}{c}P(Y\leq t+a, Y>a|A=a)da\\
&=&\int_0^c \frac{1}{c}P(a<Y\leq t+a)da\\
&=&c^{-1}\int_0^c \{F(t+a)-F(a)\}da,
\end{eqnarray*}
where the second last step follows from the assumption that $A$ and $Y$ are independent.
Similarly,
\begin{eqnarray*}
P(Y>A)&=&\int_0^c \frac{1}{c}P(Y>a|A=a)da\\
&=&\int_0^c \frac{1}{c}P(Y>a)da\\
&=&c^{-1}\int_0^c\{1-F(a)\}da.
\end{eqnarray*}
As a consequence,
\[
P(V\leq t|Y>A)
=\frac{\int_0^c \{F(t+a)-F(a)\}da}{\int_0^{c} \{1-F(a)\} da}.
\]
Hence, the pdf of $V$ conditional on $Y>A$ is
$$
\frac{\int_0^c f(t+a)da}{\int_0^{c}\{1-F(a)\} da}
=\frac{1-F(t)}{\int_0^{c}\{1-F(a)\}da}.
$$

If we let $c\rightarrow \infty$, the pdf of $V$ conditional on $Y>A$ becomes
\[
 \frac{1-F(t) }{\int_0^{\infty}\{1-F(a)\} da},
\]
which becomes precisely the forward time distribution in the renewal process when reaching equilibrium status.
It can be verified that
$$
\int_0^{\infty}\{1-F(a)\} da=\int_{0}^\infty a f(a) da.
$$
Therefore, the pdf of $V$  conditional on $Y>A$ is
$$
g(t)
= \frac{ {\int_{t}^\infty f(a) da}} {\int_{0}^\infty a f(a) da}
\quad
\mbox{for}
~ t>0.
$$

Note that, even if $Y$ has a finite support, we may choose a large $c$ such that $F(t)=$1 for $t\geq c$.
Also, note that the uniform assumption of $A$ is very common, since in the early outbreak stage we expect the number of people departing Wuhan everyday should be uniformly distributed.
Lastly, regarding the equilibrium assumption in our real data, the sample size of 1211 cases were collected as of February 15, 2020, and their travel data of leaving Wuhan were between January 19 and January 23.
This enabled us to have an average follow-up time for symptoms onset of as long as 25 days.
With an adequate long run, the renewal process would reach the equilibrium status.
In summary, the forward time distribution $g(t)$ in the renewal process is a good approximation to the truncation distribution of $V$.

\section{Goodness-of-fit Test of Model (1.2) \label{S2}}

In this section, we first review the goodness-of-fit test of \cite{Deng2020} for model (1.2) in the main paper,
and then apply it to check whether model (1.2) is suitable for the data analyzed in Section 5 of the main paper.

Recall that model (1.2) posited that $t_1,\ldots,t_n$ are $n$ iid observations from
\begin{equation*}
h(t;\lambda,\alpha,p)= p f(t;\lambda,\alpha)+(1-p) g(t;\lambda,\alpha),\quad t>0,
\end{equation*}
with $f(t;\lambda,\alpha)$ being the pdf of a pre-specified distribution such as a Weibull distribution, and
$g(t;\lambda,\alpha)= \frac{ {\int_{t}^\infty f(y;\lambda,\alpha) dy}} {\int_{0}^\infty y f(y;\lambda,\alpha) dy}$ being a biased sampling version of $f(t;\lambda,\alpha)$.
The idea of this test is to divide the non-negative real line into $k$ disjoint and adjacent intervals, whereupon the goodness-of-fit statistic is defined as
\[
G_n =\sum_{i=1}^{k}\frac{(O_i-E_i)^2}{E_i},
\]
where $O_i$ is the observed number of cases in the $i$th interval, $E_i$ is the expected number of cases in the $i$th interval based on $h(t;\hat\lambda,\hat\alpha,\hat p)$, and $k$ is chosen such that $E_i\geq 5$ for each interval. The asymptotic null distribution of $G_n$ is known to be a chi-squared distribution with $k-3-1$ degrees of freedom because there are three parameters in total in model (1.2).

For the data in Section~5 of the main paper, \cite{Deng2020} first partitioned the non-negative real line into $k=17$ intervals: [0,0.5), $[i-0.5,i+0.5)$ for $i=1,\ldots,15$, and $[15.5,\infty)$. When $f(t;\lambda,\alpha)$ is the pdf of a Weibull distribution, the observed value of $G_n$ is 14.09 with an asymptotic $p$-value of 0.37, calibrated by the $\chi^2_{13}$ distribution. Hence, we do not have strong evidence for rejecting model (1.2) with $f(t;\lambda,\alpha)$ being the pdf of a Weibull distribution for the duration-time data in Section~5 of the main paper.

\section{Regularity Conditions \label{S3}}

Our asymptotic results about $R_n$ in Theorems 2-4 rely on the
following regularity conditions, in which the expectation is taken with respect to the null model.
\begin{enumerate}
\item[]C1. (i) For sufficiently small $\epsilon>0$, $\e[ \log \{ 1+ f_\epsilon(T) \} ]<\infty$ and $\e[ \log \{ 1+ g_\epsilon(T)\} ]<\infty$, where $ f_\epsilon(t)= \sup _{(\lambda-\lambda_0)^2+(\alpha-\alpha_0)^2<\epsilon^2} f(t; \lambda,\alpha)$ and $g_\epsilon(t)$ is similarly defined; (ii) for sufficiently large $r>0$, $\e[ \log\{ 1+ \varphi_{f,r}(T) \}]<\infty$ and $\e[ \log\{ 1+ \varphi_{g,r}(T)\} ]<\infty$, where $\varphi_{f,r}(t) =\sup _{\lambda^2+\alpha^2\geq r^2} f(t; \lambda,\alpha)$ and $\varphi_{g,r}(t) $ is similarly defined; (iii) $f(t; \lambda,\alpha) \to 0$ and $g(t; \lambda,\alpha) \to 0$ as $ \lambda^2+\alpha^2\to \infty$.
\item[]C2. The parameters $\lambda$ and $\alpha$ are identifiable.
\item[]C3. $f(t;\lambda,\alpha)$ has common support and continuous third-order partial derivatives with respect to $\lambda$ and $\alpha$.
\item[]C4. ${\bf B}$ is positive definite.
\item[]C5. For two non-negative integers $h$ and $l$ such that $h+l\leq 2$, there exists a function $G(t)$ with $\e\{G(T)\}<\infty$ such that
\[
\Big|
\frac{ \partial^{h+l} f(t; \lambda_0,\alpha_0)/\partial \lambda^h \partial \alpha^l } { f(t; \lambda_0,\alpha_0)}
\Big|^3\leq G(t)
~~
\mbox{and}
~~
\Big|
\frac{ \partial^{h+l} g(t; \lambda_0,\alpha_0)/\partial \lambda^h \partial \alpha^l } { g(t; \lambda_0,\alpha_0)}
\Big|^3\leq G(t).
\]
Moreover, there exists a positive $\epsilon_0$ such that for $h+l=3$,
\[
\sup_{ (\lambda-\lambda_0)^2+(\alpha-\alpha_0)^2\leq \epsilon^2_0}
\Big|
\frac{ \partial^{h+l} f(t; \lambda,\alpha)/\partial \lambda^h \partial \alpha^l } { f(t; \lambda_0,\alpha_0)}
\Big|^3\leq G(t)
\]
and
\[
\sup_{ (\lambda-\lambda_0)^2+(\alpha-\alpha_0)^2\leq \epsilon_0^2}
\Big|
\frac{ \partial^{h+l} g(t; \lambda,\alpha)/\partial \lambda^h \partial \alpha^l } { g(t; \lambda_0,\alpha_0)}
\Big|^3\leq G(t).
\]
\end{enumerate}

\section{Proof of Theorem~1 \label{S4}}
Recall that $F(t;\lambda,\alpha)$ is the cumulative distribution function corresponding to $f(t;\lambda,\alpha)$
and
$$
h(t;\lambda,\alpha,p)= p f(t;\lambda,\alpha)+(1-p) g(t;\lambda,\alpha),\quad t>0,
$$
where
$$
g(t;\lambda,\alpha)=\frac{1-F(t;\lambda,\alpha)}{\mu(\lambda,\alpha)}~~
\mbox{ with }
~~
\mu(\lambda,\alpha)=\int_{0}^\infty t f(t;\lambda,\alpha) dt.
$$
Then $h(t;\lambda,\alpha,p)$ can be rewritten as
$$
h(t;\lambda,\alpha,p)= p f(t;\lambda,\alpha)+(1-p) \frac{1-F(t;\lambda,\alpha)}{\mu(\lambda,\alpha)},\quad t>0.
$$

{\bf For (a)}. We concentrate on the case in which
\begin{equation}
\label{iden0}
A(\lambda_1,\alpha_1)=\lim_{t\to \infty}\frac{f(t;\lambda_1,\alpha_1)}{1-F(t;\lambda_1,\alpha_1)}=0.
\end{equation}
The proof for the case in which $
A(\lambda_1,\alpha_1)=\infty
$
is similar.

We first argue that $(\lambda_1,\alpha_1)=(\lambda_2,\alpha_2)$ when
$h(t;\lambda_1,\alpha_1,p_1)=h(t;\lambda_2,\alpha_2,p_2)$ for all $t>0$.

If $(\lambda_1,\alpha_1)\neq(\lambda_2,\alpha_2)$, then using Condition~A2 and L'Hospital's rule,
we have
\begin{equation}
\label{iden1}
\lim_{t\to \infty} \frac{ 1-F(t;\lambda_1\alpha_1) }{ 1-F(t;\lambda_2,\alpha_2) }
=
\lim_{t\to \infty}\frac{f(t;\lambda_1,\alpha_1)}{f(t;\lambda_2,\alpha_2)}
=0
~~
\mbox{or}
~~
\infty.
\end{equation}
We further consider two different scenarios: $p_1=1$ and $p_1\neq1$.

\noindent{\it Scenario I}: $p_1\neq 1$.

Dividing $1-F(t;\lambda_1,\alpha_1)$ on both sides of
$h(t;\lambda_1,\alpha_1,p_1)=h(t;\lambda_2,\alpha_2,p_2)$,
we have
\begin{equation}
\label{iden2}
\frac{p_1 f(t;\lambda_1,\alpha_1)}{1-F(t;\lambda_1,\alpha_1)}
+\frac{(1-p_1)}{\mu(\lambda_1,\alpha_1)}
=
\frac{p_2 f(t;\lambda_2,\alpha_2)}{1-F(t;\lambda_1,\alpha_1)}
+ \frac{(1-p_2)\{ 1-F(t;\lambda_2,\alpha_2)\}}{\mu(\lambda_2,\alpha_2)\{1-F(t;\lambda_1,\alpha_1)\}}.
\end{equation}
When $t\to\infty$ in \eqref{iden2}, by \eqref{iden0}--\eqref{iden1} and Condition~A3,
the left-hand side becomes a positive number $ (1-p_1)/\mu(\lambda_1,\alpha_1) $,
whereas the right-hand side becomes either 0 or $\infty$,
which is a contradiction.

\noindent{\it Scenario II}: $p_1=1$.

When $p_1=1$, $h(t;\lambda_1,\alpha_1,p_1)=h(t;\lambda_2,\alpha_2,p_2)$ implies that
$$
f(t;\lambda_1,\alpha_1)
=
p_2 f(t;\lambda_2,\alpha_2)
+ \frac{(1-p_2)\{ 1-F(t;\lambda_2,\alpha_2)\}}{\mu(\lambda_2,\alpha_2)}.
$$
Dividing $f(t;\lambda_1,\alpha_1)$ on both sides of
the above equation gives
\begin{equation}
\label{iden3}
1
=
\frac{p_2 f(t;\lambda_2,\alpha_2)}{f(t;\lambda_1,\alpha_1)}
+ \frac{(1-p_2)\{ 1-F(t;\lambda_2,\alpha_2)\}}{\mu(\lambda_2,\alpha_2) f(t;\lambda_1,\alpha_1) }.
\end{equation}
When $t\to\infty$ in \eqref{iden3}, by Conditions~A2 and A3,
the right-hand side is equal to either 0 or $\infty$,
whereas the left-hand side is equal to 1,
which is a contradiction.

In summary, if $h(t;\lambda_1,\alpha_1,p_1)=h(t;\lambda_2,\alpha_2,p_2)$ for all $t>0$
and $A(\lambda_1,\alpha_1)=0$,
then under Conditions~A1--A3, we must have
$$
(\lambda_1,\alpha_1)=(\lambda_2,\alpha_2).
$$
This, together with $h(t;\lambda_1,\alpha_1,p_1)=h(t;\lambda_2,\alpha_2,p_2)$, implies that
$$
p_1-p_2=
(p_1-p_2)\frac{ \mu(\lambda_1,\alpha_1) f(t;\lambda_1,\alpha_1)}{ 1-F(t;\lambda_1,\alpha_1)}
$$
for all $t>0$.
Letting $t\to\infty$ in the above equation and noting that $A(\lambda_1,\alpha_1)=0$, we obtain
$
p_1=p_2.
$
Hence $(\lambda_1,\alpha_1,p_1)=(\lambda_2,\alpha_2,p_2)$, as claimed in (a).

{\bf For (b).}
We first argue that $(\lambda_1,\alpha_1)=(\lambda_2,\alpha_2)$ when
$0<A(\lambda_1,\alpha_1)<\infty$
and
$h(t;\lambda_1,\alpha_1,p_1)=h(t;\lambda_2,\alpha_2,p_2)$ for all $t>0$.

If $(\lambda_1,\alpha_1)\neq(\lambda_2,\alpha_2)$,
when $t\to\infty$ in \eqref{iden2}, by \eqref{iden1} and Condition~A3,
the left-hand side of \eqref{iden2} becomes
$
p_1 A(\lambda_1,\alpha_1)
+\frac{(1-p_1)}{\mu(\lambda_1,\alpha_1)},
$
which is finite and positive,
while the right-hand side of \eqref{iden2} is equal to either 0 or $\infty$,
which is a contradiction.
Hence we must have $(\lambda_1,\alpha_1)=(\lambda_2,\alpha_2)$.
This completes the first part of (b).

Recall that $(\lambda_1,\alpha_1)=(\lambda_2,\alpha_2)$ implies that
$$
p_1-p_2=
(p_1-p_2)\frac{ \mu(\lambda_1,\alpha_1) f(t;\lambda_1,\alpha_1)}{ 1-F(t;\lambda_1,\alpha_1)}
$$
for all $t>0$.
If $\frac{f(t;\lambda_1,\alpha_1)}{1-F(t;\lambda_1,\alpha_1)}$ is not a constant function of $t$, then we must have $p_1=p_2$.
If $\frac{f(t;\lambda_1,\alpha_1)}{1-F(t;\lambda_1,\alpha_1)}$ is a constant function of $t$,
then $\frac{ \mu(\lambda_1,\alpha_1) f(t;\lambda_1,\alpha_1)}{ 1-F(t;\lambda_1,\alpha_1)}$
must equal 1 for all $t>0$
because both $f(t;\lambda_1,\alpha_1)$ and
$\frac{ 1-F(t;\lambda_1,\alpha_1)}{ \mu(\lambda_1,\alpha_1)}$
are probability density functions.
In this case, $p_1$ and $p_2$ need not be equal.
This completes the second part of (b).

\section{Proof of Theorem~2}

\subsection{Two technical lemmas}
We first establish two technical lemmas.
Lemma~\ref{lem1} establishes the consistency
of the maximum likelihood estimator (MLE) under the null model; this is the first step in the proof of Theorem~2.
The lemma claims that any estimator of $(\lambda, \alpha, p)$
with a large likelihood value is consistent for $\lambda$ and $\alpha$
under the null model.
Recall that the true values of $\lambda$ and $\alpha$ under the null model are
$\lambda_0$ and $\alpha_0$, respectively.

\begin{lemma}
\label{lem1}
Assume the conditions of Theorem~2.
Let $(\bar\lambda, \bar\alpha, \bar p)$ be any estimator of
$(\lambda, \alpha, p)$ such that
\begin{eqnarray}
\label{cond.lem1}
l_n(\bar\lambda, \bar\alpha, \bar p)-l_n(\lambda_0,\alpha_0,1)>c
>-\infty
\end{eqnarray}
for some constant $c$ for all $n$.
Then under the null model, $\bar\lambda-\lambda_0=o_p(1)$ and $\bar\alpha-\alpha_0=o_p(1)$.
\end{lemma}

\proof
Under Condition~C2, both $\lambda$ and $\alpha$ are identifiable under the null hypothesis, although $p$ is not.
The proof then follows by using techniques similar to those in Lemma~1 of \cite{Li2009}
and \cite{Wald1949}.
\qed

In the next lemma, we strengthen the conclusion of Lemma~\ref{lem1} by providing an order assessment of the estimators. Recall that
\begin{eqnarray*}
X_i&=&\frac{\partial f(t_i;\lambda_0,\alpha_0)/\partial \lambda}{f(t_i; \lambda_0,\alpha_0)},\\
Y_{i1}&=&\frac{\partial f(t_i;\lambda_0,\alpha_0)/\partial \alpha}{f(t_i; \lambda_0,\alpha_0)},
\label{Yi1}\\
Y_{i2}&=&\frac{\partial g(t_i;\lambda_0,\alpha_0)/\partial \alpha}{g(t_i; \lambda_0,\alpha_0)}.
\label{Yi2}
\end{eqnarray*}
Note that under Condition~C0,
$$
h(t_i;\lambda_0,\alpha_0,1)
=
f(t_i;\lambda_0,\alpha_0)=g(t_i;\lambda_0,\alpha_0)
~~
\mbox{ and }
~~
\frac{\partial g(t_i;\lambda_0,\alpha_0)/\partial \lambda}{g(t_i; \lambda_0,\alpha_0)} =X_i.
$$
Define ${\bf b}_i=(X_i,Y_{i1},Y_{i2})^\T$.
Then $\e({\bf b}_i)={\bf 0}$
and we denote the variance-covariance matrix
\begin{equation}
{\bf B}=\var({\bf b}_i)
=\left(
\begin{array}{ccc}
B_{11}&B_{12}&B_{13}\\
B_{21}&B_{22}&B_{23}\\
B_{31}&B_{32}&B_{33}\\
\end{array}
\right),
\end{equation}
where the expectation and variance are taken with respect to the null model $f(t;\lambda_0,\alpha_0)$.
\begin{lemma}
\label{lem2}
Assume the conditions of Lemma~\ref{lem1}.
Then under the null model, $\bar\lambda-\lambda_0=O_p(n^{-1/2})$ and $\bar\alpha-\alpha_0=O_p(n^{-1/2})$.
\end{lemma}
\proof
In the following, we will first derive an upper bound for $\ell_{n}(\bar \lambda, \bar \alpha,\bar p)-\ell_{n}(\alpha_0,\lambda_0,1)
$.
Then together with the lower bound $c$, we obtain the order assessment of $\bar\lambda$ and $\bar\alpha$.
Write
$$
\ell_{n}(\bar \lambda, \bar \alpha,\bar p)-\ell_{n}(\alpha_0,\lambda_0,1)
=\sum_{i=1}^{n}\log\{1+\delta_i(\bar \lambda, \bar \alpha,\bar p) \}
$$
with
\begin{eqnarray*}
\delta_i(\bar \lambda, \bar \alpha,\bar p)
&=&\frac{\bar p f(t_i;\bar\lambda,\bar\alpha)+(1-\bar p) g(t_i; \bar\lambda,\bar\alpha)}
{h(t_i;\alpha_0,\lambda_0,1)}-1\\
&=&\bar p \frac{f(t_i;\bar\lambda,\bar\alpha)- f(t_i;\lambda_0,\alpha_0)}{f(t_i;\lambda_0,\alpha_0)}+(1-\bar p)\frac{g(t_i;\bar\lambda,\bar\alpha)- g(t_i;\lambda_0,\alpha_0)}{g(t_i;\lambda_0,\alpha_0)}.
\end{eqnarray*}
By the inequality $\log(1+x)\leq x-x^2/2+x^3/3$, we have
\begin{equation}
\label{ln.upper1}
\ell_{n}(\bar \lambda, \bar \alpha,\bar p)-\ell_{n}(\alpha_0,\lambda_0,1)
\leq \sum_{i=1}^{n}\delta_i(\bar \lambda, \bar \alpha,\bar p)
-\sum_{i=1}^{n}\delta_i^2(\bar \lambda, \bar \alpha,\bar p)/2+
\sum_{i=1}^{n}\delta_i^3(\bar \lambda, \bar \alpha,\bar p)/3.
\end{equation}

From Lemma~\ref{lem1}, we have the consistency results $\bar\lambda-\lambda_0=o_p(1)$ and $\bar\alpha-\alpha_0=o_p(1)$.
Applying a first-order Taylor expansion to $f(t_i;\bar\lambda,\bar\alpha)$ and $g(t_i;\bar\lambda,\bar\alpha)$,
we find that
$$
\delta_i(\bar \lambda, \bar \alpha,\bar p)
=
(\bar\lambda-\lambda_0) X_i+\bar p(\bar\alpha-\alpha_0) Y_{i1} +(1-\bar p)(\bar\alpha-\alpha_0) Y_{i2}+ \varepsilon_{in},
$$
and
the remainder term $\varepsilon_n=\sum_{i=1}^{n}\varepsilon_{in}$
satisfies
\begin{equation*}
\varepsilon_n=O_p(n^{1/2})\left\{(\bar\lambda-\lambda_0) ^2+(\bar\alpha-\alpha_0) ^2 \right\}.
\end{equation*}

Let $\bar s_1=\bar\lambda-\lambda_0$, $\bar s_2=\bar p(\bar\alpha-\alpha_0)$, $\bar s_3=(1-\bar p)(\bar\alpha-\alpha_0)$, and $\bar{\bf s}=(\bar s_1,\bar s_2,\bar s_3)^\T$.
Then
$$
\delta_i(\bar \lambda, \bar \alpha,\bar p)
=\bar {\bf s}^\T {\bf b}_i + \varepsilon_{in}
$$
and
\begin{equation}
\label{error1}
\varepsilon_n=O_p(n^{1/2})\bar {\bf s}^\T \bar {\bf s}=o_p(n)\bar {\bf s}^\T \bar {\bf s} .
\end{equation}
Therefore, for the linear term in \eqref{ln.upper1},
we have
\begin{equation}
\label{ln1}
\sum_{i=1}^{n}\delta_i(\bar \lambda, \bar \alpha,\bar p)=
\bar {\bf s}^\T \sum_{i=1}^n{\bf b}_i
+\varepsilon_n,
\end{equation}
where the order of $\varepsilon_n$ is assessed in \eqref{error1}.

After some work, we can further show that
\begin{eqnarray*}
\label{ln2}
\sum_{i=1}^{n}\delta_i^2(\bar \lambda, \bar \alpha,\bar p)&=&
\sum_{i=1}^n\left(\bar {\bf s}^\T {\bf b}_i\right)^2 +O_p(\varepsilon_n),\\
\label{ln3}
\sum_{i=1}^{n}\delta_i^3(\bar \lambda, \bar \alpha,\bar p)&=&
\sum_{i=1}^n\left(\bar {\bf s}^\T {\bf b}_i\right)^3 +O_p(\varepsilon_n).
\end{eqnarray*}
By the strong law of large numbers and Condition~C4 that ${\bf B}$ is positive definite,
we further have
\begin{eqnarray}
\label{ln2}
\sum_{i=1}^{n}\delta_i^2(\bar \lambda, \bar \alpha,\bar p)&=&
n \bar {\bf s}^\T{\bf B}\bar {\bf s}+o_p(n)\bar {\bf s}^\T \bar {\bf s},\\
\label{ln3}
\sum_{i=1}^{n}\delta_i^3(\bar \lambda, \bar \alpha,\bar p)&=&o_p(n)\bar {\bf s}^\T \bar {\bf s}.
\end{eqnarray}

Combining \eqref{ln.upper1}--\eqref{ln3}, we obtain the refined upper bound for $\ell_{n}(\bar \lambda, \bar \alpha,\bar p)-\ell_{n}(\alpha_0,\lambda_0,1)$ as follows:
\begin{eqnarray}
\label{ln.upper2}
\ell_{n}(\bar \lambda, \bar \alpha,\bar p)-\ell_{n}(\alpha_0,\lambda_0,1)
\leq
\bar {\bf s}^\T \sum_{i=1}^n{\bf b}_i
-0.5 n \bar {\bf s}^\T{\bf B}\bar {\bf s} \{1+o_p(1)\}.
\end{eqnarray}
Because ${\bf B}$ is positive definite, the upper bound in \eqref{ln.upper2}
is of order $O_p(1)$.
Together with the lower bound $c$, this implies that
$$
\bar {\bf s}=O_p(n^{-1/2}).
$$
Any values of $\bar {\bf s}$ outside
this range will violate
the inequality.
Note that $\bar {\bf s}$ implies that $\bar\lambda-\lambda_0=O_p(n^{-1/2})$ and $\bar\alpha-\alpha_0=O_p(n^{-1/2})$.
This completes the proof.
\qed

\subsection{Proof of Theorem~2}
Note that
\begin{eqnarray}
\label{LLR}
R_n=2\left\{ \ell_n(\hat\lambda,\hat\alpha,\hat p)-\ell_n(\hat\lambda_0,\alpha_0,1)\right\}=R_{1n}-R_{2n},
\end{eqnarray}
where $$R_{1n}=2\left\{ \ell_n(\hat\lambda,\hat\alpha,\hat p)-\ell_{n}(\lambda_0,\alpha_0,1) \right\},
~~R_{2n}=2\left\{\ell_n(\hat\lambda_0,\alpha_0,1)-\ell_{n}(\lambda_0,\alpha_0,1)\right\}. $$

Applying some of the classical results for regular models \citep{Serfling1980}, we have
\begin{equation}
\label{LLR2}
R_{2n}=\frac{\left(n^{-1/2}\sum_{i=1}^nX_i\right)^2}{B_{11}}+o_p(1).
\end{equation}

Next, we use a sandwich method to find the approximation of $R_{1n}$.
We proceed in two steps. In step~1, we derive an upper bound for $R_{1n}$
and in step~2, we argue that the upper bound is achievable.

Let
$
(\hat\lambda_p,\hat\alpha_p)=\arg\max_{\lambda,p} \ell_{n}(\lambda, \alpha, p)
$
be the constrained MLE of $(\lambda,p)$ for given $p$.
Define
$
R_{1n}(p)=2\left\{ \ell_n(\hat\lambda_p,\hat\alpha_p, p)-\ell_{n}(\lambda_0,\alpha_0,1) \right\}.
$
Then $R_{1n} =\sup_p R_{1n}(p)$.
{By the definition of $(\hat\lambda_p,\hat\alpha_p)$,
we have $\ell_n(\hat\lambda_p,\hat\alpha_p, p)-\ell_{n}(\lambda_0,\alpha_0,1) \geq0$.
Hence, Condition~\eqref{cond.lem1} is satisfied.
Then applying the results in Lemma~\ref{lem2} and \eqref{ln.upper2}, we obtain
$$
R_{1n}(p)
\leq
2\hat {\bf s}^\T (p) \sum_{i=1}^n{\bf b}_i
-n \hat {\bf s}^\T (p) {\bf B}\hat {\bf s} (p)+o_p(1),
$$
where
$
\hat {\bf s} (p)$ is defined similarly to $\bar {\bf s} $ with
$(\hat\lambda_p, \hat \alpha_p, p)$
in place of $(\bar\lambda, \bar \alpha,\bar p)$.

Define
$$
\hat{\bf t}(p)=\Big(\hat t_1(p),\hat t_2(p) \Big)^\T=\Big(\hat\lambda_p-\lambda_0, \hat\alpha_p-\alpha_0\Big)^\T,
~~{\bf c}_i(p)=\Big(X_i,Y_i(p)\Big)^\T
$$
with
$
Y_i(p)=pY_{i1}+(1-p)Y_{i2}
$, and ${\bf C}(p)= \var\{{\bf c}_i(p) \}$.
Then after some algebra, we obtain a refined upper bound for $R_{1n}(p)$ as
\begin{equation}
\label{r1n.upper1}
R_{1n}(p)
\leq
2\hat{\bf t}^\T (p) \sum_{i=1}^n{\bf c}_i(p)
-n \hat {\bf t}^\T (p) {\bf C}(p)\hat {\bf t} (p)+o_p(1).
\end{equation}

To further simplify the upper bound in \eqref{r1n.upper1},
let
$$
a(p)=p\frac{B_{12}}{B_{11}}+(1-p)\frac{B_{13}}{B_{11}},~~
\hat t_1^*(p)=\hat\lambda_p-\lambda_0+a(p)(\hat\alpha_p-\alpha_0),
$$
and
\begin{equation}
\label{def.zi}
Z_i(p)=Y_i(p)-a(p)X_i.
\end{equation}
It can be verified that $\cov\left\{ X_i, Z_i(p)\right\}=0$
and
$\var\left\{ Z_i(p)\right\}=\sigma(p,p)$, where $\sigma(\cdot,\cdot)$
is defined in (3.3) of the main paper.
Then
the upper bound in \eqref{r1n.upper1} becomes
\begin{eqnarray}
\nonumber
\hspace{-0.2in}R_{1n}(p)
&\leq&
2\hat t_1^*(p) \sum_{i=1}^n X_i-nB_{11} \{ \hat t_1^*(p) \}^2\\
&&
+ 2\hat t_2(p) \sum_{i=1}^n Z_i(p)- n\sigma(p,p)\{ \hat t_2(p) \}^2 +o_p(1)\nonumber\\
&\leq&\frac{\left(n^{-1/2}\sum_{i=1}^nX_i\right)^2}{B_{11}}+
\left\{\frac{n^{-1/2}\sum_{i=1}^n Z_i(p)}{\sqrt{\sigma(p,p)}}\right\}^2+o_p(1).
\label{r1n.upper2}
\end{eqnarray}

Next, we show that the upper bound in \eqref{r1n.upper2} for $R_{1n}(p)$ is
achievable.
Let
$(\tilde\lambda_p,\tilde\alpha_p)$
be determined by
$$
\tilde\lambda_p-\lambda_0+ a(p) (\tilde \alpha_p-\alpha_0)
=n^{-1/2}\sum_{i=1}^n X_i/B_{11},
\quad
\tilde \alpha_p-\alpha_0= \frac{n^{-1/2}\sum_{i=1}^n Z_i( p)}{\sqrt{\sigma( p, p)}}.
$$
Note that it is easy to verify that $(\tilde\lambda_p,\tilde\alpha_p)$ exists and
$$
\tilde\lambda_p-\lambda_0=O_p(n^{-1/2}),
\quad
\tilde \alpha_p-\alpha_0=O_p(n^{-1/2})
$$
uniformly over $p$.
With this order assessment and applying a second-order Taylor expansion, we
have
\begin{eqnarray}
\nonumber
\hspace{-0.2in}
R_{1n}(p)&\geq&2\left\{\ell_n(\tilde\lambda_p,\tilde\alpha_p, p)
-\ell_n(\lambda_0,\alpha_0,1)\right\}\\
&=& \frac{\left(n^{-1/2}\sum_{i=1}^nX_i\right)^2}{B_{11}}+
\left\{\frac{n^{-1/2}\sum_{i=1}^n Z_i(p)}{\sqrt{\sigma(p,p)}}\right\}^2+o_p(1).
\label{r1n.lower}
\end{eqnarray}
Combining \eqref{r1n.upper2} and \eqref{r1n.lower} leads to
$$
R_{1n}(p) = \frac{\left(n^{-1/2}\sum_{i=1}^nX_i\right)^2}{B_{11}}+
\left\{\frac{n^{-1/2}\sum_{i=1}^n Z_i(p)}{\sqrt{\sigma(p,p)}}\right\}^2+o_p(1).
$$
Hence
\begin{equation}
\label{LLR1}
R_{1n}=\sup_p R_{1n}(p)= \frac{\left(n^{-1/2}\sum_{i=1}^nX_i\right)^2}{B_{11}}+
\sup_p\left\{\frac{n^{-1/2}\sum_{i=1}^n Z_i(p)}{\sqrt{\sigma(p,p)}}\right\}^2+o_p(1),
\end{equation}
which together with \eqref{LLR2} gives
$$
R_n= \sup_p\left\{\frac{n^{-1/2}\sum_{i=1}^n Z_i(p)}{\sqrt{\sigma(p,p)}}\right\}^2+o_p(1).
$$

Recall the form of $Z_i(p)$ in \eqref{def.zi}.
We can rewrite it as
$$
Z_i(p)=
Z_{1i}+pZ_{2i}
$$
with $Z_{i1}=Y_{i2}-(B_{13}/B_{11})X_i $ and
$$Z_{i2}=\{Y_{i1}-(B_{12}/B_{11})X_i\}- \{Y_{i2}-(B_{13}/B_{11})X_i\} .
$$
It can be verified that
$
\e(Z_{i1})=\e(Z_{i2})=0$
and
$$
\var(Z_{1i})=\sigma_{11},\quad \var(Z_{2i})=\sigma_{22},\quad \cov(Z_{1i},Z_{2i})=\sigma_{12}.
$$
Hence
$$
R_n= \sup_p\left\{\frac{n^{-1/2}\sum_{i=1}^n Z_i(p)}{\sqrt{\sigma(p,p)}}\right\}^2+o_p(1)
\stackrel{d}\rightarrow R=\sup_{p}Z^2(p),
$$
where
$
Z(p)={(Z_1+pZ_2)}/{\sqrt{\sigma(p,p)}}
$
with
\begin{equation}
\label{dis.z}
(Z_1,Z_2)^\T
\sim N\left(
\left(
\begin{array}{c}
0\\
0\\
\end{array}
\right),
\left(
\begin{array}{cc}
\sigma_{11}&\sigma_{12}\\
\sigma_{12}&\sigma_{22}\\
\end{array}
\right)
\right).
\end{equation}
It can be verified that the process
$Z(p)$ is a Gaussian process with zero mean,
unit variance, and covariance function
\begin{equation*}
\cov\big\{Z(p_1), Z(p_2)\big\}
=\frac{\sigma(p_1,p_2)}{\sqrt{\sigma(p_1,p_1) \sigma(p_2,p_2) }}.
\end{equation*}
This completes the proof.
\qed

\section{Proof of Theorem~3}
Recall that $Z(p)={(Z_1+pZ_2)}/{\sqrt{\sigma(p,p)}}$ with the joint distribution of $(Z_1,Z_2)^\T$
provided in \eqref{dis.z}.
Let
$$
W_1=\left(Z_1-\frac{\sigma_{12}}{\sigma_{22}}Z_2\right)/a_1,
\quad
W_2=Z_2/a_2,
$$
where
$$
a_1=\sqrt{\sigma_{11}-\frac{\sigma_{12}^2}{\sigma_{22}}},
\quad
a_2=\sqrt{\sigma_{22}}.
$$
By construction, it can be verified that $W_1$ and $W_2$ are two independent $N(0,1)$ random variables,
and
$$
Z(p)=\frac{a_1W_1+\left(p+\frac{\sigma_{12}}{\sigma_{22}}\right) a_2 W_2}{\sqrt{ \sigma(p,p) }}.
$$

To find a simpler form for $Z(p)$, we consider two polar transformations.
The first one is defined in the main paper:
$$
(\cos\theta, \sin\theta)=\big(c_1(p),c_2(p)\big),
$$
where
$$
c_1(p)=\frac{a_1}{\sqrt{\sigma(p,p)}}
~~
\mbox{ and }
~~
c_2(p)=\frac{(p+\sigma_{12}/\sigma_{22})a_2}{\sqrt{\sigma(p,p)}}.
$$
By Condition~C6, we have
$$
\left\{\big(c_1(p),c_2(p)\big):0\leq p \leq 1\right\}=\{(\cos\theta,\sin\theta):
\Delta_1\leq\theta\leq\Delta_2\}.
$$
The second polar transformation is
$$
(W_1,W_2)=(\rho\cos\eta,\rho\sin\eta),
$$
where $\rho^2$ with $\rho>0$ and $\eta$ are two independent
random variables with $\rho^2$ from a $\chi_2^2$ distribution and
$\eta$ from a uniform distribution on $[-\pi, \pi]$.
Then
$$
Z(p)=\rho\cos\eta \cos\theta+\rho\sin\eta \sin\theta=\rho\cos(\theta-\eta)
$$
and
$$
\sup_p Z^2(p)=\sup_{\theta\in[\Delta_1,\Delta_2]} \rho^2\cos^2(\theta-\eta).
$$
After some algebra, we can check that
$$
\sup_{\theta\in[\Delta_1,\Delta_2]} \rho^2\cos^2(\theta-\eta)
=
\rho^2 \{ I (\eta\in A_1) + I (\eta\in A_2) \cos^2(\eta-\Delta_2) + I (\eta\in A_3) \cos^2(\eta-\Delta_1) \}.
$$
This completes the proof.
\qed

\section{Proof of Theorem~4}
We proceed in two steps.
In the first step, we show that the models under the local alternatives
\begin{equation}
\label{local.alt}
H_a^n: \lambda=\lambda_0, p=p_0, \alpha=\alpha_0+\delta n^{-1/2}
\end{equation}
are contiguous to the null model \citep{LeCam1953}.
In the second step, we find the asymptotic distribution of $R_n$ under $H_a^n$ by using Le Cam's first and third lemmas \citep{van1998}.

Let
$$
\Lambda_n=\ell_n(\lambda_0,\alpha,p_0)-\ell_n(\lambda_0,\alpha_0,1).
$$
Using the second-order Taylor expansion, under the null model, we have
$$
\Lambda_n=\sum_{i=1}^n Y_{i}(p_0)(\delta n^{-1/2}) - \frac{1}{2}\delta^2 \var\{Y_i(p_0)\}+o_p(1).
$$
By the central limit theorem, we have
$$
\Lambda_n\to N(-0.5 {d_0^2},d_0^2)
$$
in distribution under the null model, where 
$d_0=\delta^2\var\{Y_i(p_0)\}$.
Therefore, the models under the local alternatives $H_a^n$ in (\ref{local.alt}) are contiguous to the null model \citep{LeCam1953}.
This completes step~1.

Next, we move on to step~2.
Recall that under the null model,
$$
R_n= \sup_p\left\{\frac{n^{-1/2}\sum_{i=1}^n Z_i(p)}{\sqrt{\sigma(p,p)}}\right\}^2+o_p(1).
$$
By Le Cam's contiguity theory, the limiting distribution of $R_n$ under the local alternatives $H_a^n$ is determined by the joint limiting distribution of $\{n\sigma(p,p)\}^{-1/2}\sum_{i=1}^n Z_i(p)$ and $\Lambda_n$ under the null model.

By the central limit theorem and Slutsky's theorem, the joint limiting distribution of $\{n\sigma(p,p)\}^{-1/2}\sum_{i=1}^n Z_i(p)$ and $\Lambda_n$ under the null model is bivariate normal
$$
N\left(
\left(
\begin{array}{c}
0\\
-0.5d_0^2\\
\end{array}
\right),
\left(
\begin{array}{cc}
1&\omega(p,p_0)\\
\omega(p,p_0)&d_0^2\\
\end{array}
\right)
\right),
$$
where
\begin{eqnarray*}
\omega(p,p_0)
&=&\cov\left(\{\sigma(p,p)\}^{-1/2} Z_i(p), \delta Y_{i}(p_0) \right)\\
&=&\cov\left(\{\sigma(p,p)\}^{-1/2} Z_i(p), \delta Z_{i}(p_0) \right)\\
&=&\frac{\delta\sigma(p,p_0)}{\sqrt{\sigma(p,p)}}.
\end{eqnarray*}
Note that in the second equation, we have used the fact that $\cov\left\{ Z_i(p),X_i\right\}=0$ and the definition of $Z_i(p)$ in (\ref{def.zi}).

By Le~Cam's third lemma \citep{van1998}, under the local alternatives $H_a^n$,
$$
\frac{n^{-1/2}\sum_{i=1}^n Z_i(p)}{\sqrt{\sigma(p,p)}} \to N\left(\omega(p,p_0), 1\right)
$$
in distribution,
which implies that
$$
\left\{\frac{n^{-1/2}\sum_{i=1}^n Z_i(p)}{\sqrt{\sigma(p,p)}}\right\}^2 \to
\{Z(p)+\omega(p,p_0)\}^2
$$ in distribution under $H_a^n$.

Because
$$
R_n= \sup_p\left\{\frac{n^{-1/2}\sum_{i=1}^n Z_i(p)}{\sqrt{\sigma(p,p)}}\right\}^2+o_p(1)
$$
under the null model, by applying Le Cam's first lemma \citep{van1998}, we have that
$$
R_n= \sup_p\left\{\frac{n^{-1/2}\sum_{i=1}^n Z_i(p)}{\sqrt{\sigma(p,p)}}\right\}^2+o_p(1)
$$
holds also under the local alternatives $H_a^n$.
Therefore, the asymptotic distribution of $R_n$ under the local alternatives $H_a^n$ is
$$
\sup_p\left[ \left\{Z(p)+\omega(p,p_0)\right\}^2\right].
$$
This completes the proof.
\qed

\markboth{\hfill{\footnotesize\rm Chunlin Wang, Pengfei Li, Yukun Liu, Xiao-Hua Zhou and Jing Qin} \hfill}
{\hfill {\footnotesize\rm Mixture forward incubation time epidemic model} \hfill}

\bibhang=1.7pc
\bibsep=2pt
\fontsize{9}{14pt plus.8pt minus .6pt}\selectfont
\renewcommand\bibname{\large \bf References}
\expandafter\ifx\csname
natexlab\endcsname\relax\def\natexlab#1{#1}\fi
\expandafter\ifx\csname url\endcsname\relax
  \def\url#1{\texttt{#1}}\fi
\expandafter\ifx\csname urlprefix\endcsname\relax\def\urlprefix{URL}\fi


\end{document}